\documentclass[aip,amsmath,amssymb,reprint,graphicx]{revtex4-1}
\usepackage{graphicx}
\usepackage{dcolumn}
\usepackage{bm}
\usepackage{float}

\usepackage[utf8]{inputenc}
\usepackage[T1]{fontenc}
\usepackage{mathptmx}
\usepackage{siunitx}  
\usepackage{amsmath}
\usepackage{subfigure}
\usepackage[hidelinks]{hyperref}
\usepackage[all]{hypcap}
\usepackage[noabbrev, capitalise]{cleveref}
\usepackage{amssymb}

\draft

\begin{document}

\preprint{AIP/123-QED}

\title{Rigid body dynamics of diamagnetically levitating graphite resonators} 

\author{Xianfeng Chen}
\affiliation{Department of Precision and Microsystems Engineering, Delft University of Technology, Mekelweg 2, 2628 CD, Delft, The Netherlands}

\author{Ata Keşkekler}
\affiliation{Department of Precision and Microsystems Engineering, Delft University of Technology, Mekelweg 2, 2628 CD, Delft, The Netherlands}

\author{Farbod Alijani}
\email{f.alijani@tudelft.nl}
\affiliation{Department of Precision and Microsystems Engineering, Delft University of Technology, Mekelweg 2, 2628 CD, Delft, The Netherlands}

\author{Peter G. Steeneken}
\email{p.g.steeneken@tudelft.nl}
\affiliation{Department of Precision and Microsystems Engineering, Delft University of Technology, Mekelweg 2, 2628 CD, Delft, The Netherlands}
\affiliation{Kavli Institute of Nanoscience, Delft University of Technology, Lorentzweg 1, 2628 CJ, Delft, The Netherlands.}

\begin{abstract}
Diamagnetic levitation is a promising technique for realizing resonant sensors and energy harvesters, since it offers thermal and mechanical isolation from the environment at zero power. To advance the application of diamagnetically levitating resonators, it is important to characterize their dynamics in the presence of both magnetic and gravitational fields. Here we experimentally actuate and measure rigid body modes of a diamagnetically levitating graphite plate. We numerically calculate the magnetic field and determine the influence of magnetic force on the resonance frequencies of the levitating plate. By analyzing damping mechanisms, we conclude that eddy current damping dominates dissipation in mm-sized plates. We use finite element simulations to model eddy current damping and find close agreement with experimental results. We also study the size-dependent $Q$-factors ($Q$s) of diamagnetically levitating plates and show that $Q$s above 100 million are theoretically attainable by reducing the size of the diamagnetic resonator down to microscale, making these systems of interest for next generation low-noise resonant sensors and oscillators.
\end{abstract}

\maketitle

Levitation, as the means to defy gravity, has always been a dream of mankind. This dream has been realized at large scale with the design of aircrafts, hovercrafts, and maglev trains, however, the potential of levitation is yet to be fully explored at small scale. Levitation requires forces that act in the absence of mechanical contact, like optical, acoustic, aerodynamic or magnetic forces \cite{brandt1989levitation}. Among them diamagnetic levitation stands out as the only means of stable levitation at room temperature that is passive, because it can be sustained indefinitely without feedback control or cooling systems that consume energy.  Although stable levitation in a constant magnetic field may seem to be impossible, centuries back Lord Kelvin showed that there is an exception to Earnshaw’s theorem: diamagnetic materials can levitate stably in a magnetic field \cite{kustler2007diamagnetic}. One of the most renowned experiments in this respect is the levitation of a living frog using a \SI{16.5}{Tesla} magnetic field \cite{Simon2000}. That experiment required much power to drive large electromagnets, however at small scale the story is different because the magnetic field of permanent magnets is sufficient to overcome gravitational forces and enable levitation. During the last decades, levitation of liquid droplets \cite{lyuksyutov2004chip,chetouani2006diamagnetic}, cells \cite{chetouani2007diamagnetic} and solid particles \cite{chetouani2006diamagnetic} have been demonstrated using small permanent magnets. Moreover, this passive levitation has been used for realizing devices like accelerometers \cite{garmire2007diamagnetically,pigot2009optimization}, energy harvesters \cite{liu2011nonlinear,liu2013diamagnetic,palagummi2018bi}, viscosity/density sensors \cite{clara2016advanced}, and force sensors \cite{boukallel2003levitated}. 

To realize highly sensitive resonant sensors, it is essential to characterize their frequency response and dissipation  mechanisms, that are closely linked to the precision with which frequencies of a resonant sensor can be determined and that are intrinsically coupled to the thermomechanical noise floor via the fluctuation dissipation theorem \cite{schmid2016fundamentals}. In particular, minimization of damping is an essential consideration in the design of low phase noise oscillators and highly precise resonant sensors.

One of the most important dissipation mechanisms in Micro-Electro-Mechanical Systems (MEMS) is acoustic loss, which occurs when mechanical energy leaves the structure as sound waves, via the anchors or clamping points. In literature great efforts are undertaken to minimize acoustic loss by suspending resonators via thin, high-tension tethers \cite{norte2016mechanical} or optimized clamping points \cite{van200610mhz}. However, the ultimate way of eliminating acoustic losses is levitation in vacuum, since acoustic waves cannot propagate in the absence of matter. For this reason optical levitation has been applied to obtain high $Q$ resonators \cite{gieseler2013thermal}. Even though promising, this technique has the drawback that it requires high-power lasers, that can significantly increase the temperature of the levitating object and affect its microstructure. In this respect, zero power levitation of diamagnetic materials at room temperature combined with their vacuum compatibility make them ideal candidates for tackling this challenge. Moreover, since the main sources of dissipation in diamagnetic resonators are air and eddy current damping \cite{liu2011nonlinear, palagummi2015optimal, palagummi2018bi}, by operating the resonator in vacuum, one can eliminate air damping so that the only remaining dissipative force is  eddy current force which can be minimized to obtain high $Q$ resonators.
\cite{sodano2006improved}
In this letter we study the rigid body resonances and dissipation mechanisms of a diamagnetic resonator that levitates above permanent magnets at different pressures. The diamagnetic material used in our experiments is pyrolytic graphite which has large negative magnetic susceptibility and thus can levitate easily above permanent magnets. In contrast to studies on conventional microelectromechanical resonators, these levitating resonators allow us to study the dynamics of nearly free-body objects, in the absence of mechanical clamping or anchor effects. Interestingly, the restoring force, that is generally provided by a compliant mechanical spring, is here provided by magnetic and gravitational forces. By using a laser Doppler vibrometer (LDV), 3 rigid body translational and 2 rotational modes of the plate are detected and their resonance frequencies and $Q$s are characterized experimentally. To understand the effect of air damping on our resonator, we perform experiments over a pressure range of $10^{-4}$ to \SI{1000}{mbar}. This pressure-dependent study also allows us to study the effect of eddy current damping on plates of different sizes in vacuum. To get a deeper insight into the magnetically induced stiffness and eddy current damping of the resonator, numerical finite element method (FEM) models are used to simulate the magnetic force and eddy currents on the levitating plate. With our FEM model, the effect of plate size on the $Q$ is studied, from which we conclude that the $Q$s of rigid body modes for diamagnetically levitating objects increases at small dimensions.

\begin{figure}
	\centering
	\includegraphics[width=8.5cm]{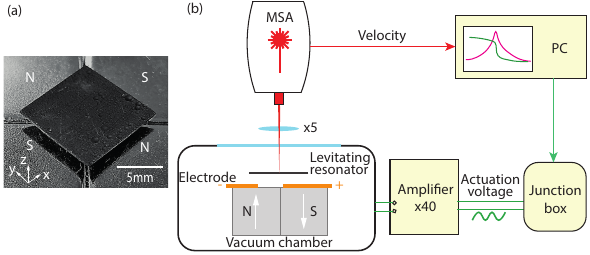}
	\caption{The levitating resonator and experimental setup.  (a) Levitating pyrolytic graphite plate above 4 cubic Nickel coated NdFeB magnets with alternating magnetization, where N and S stand for north and south pole of the magnet, respectively. The origin of the coordinate system is at the center of the plate. (b) Schematic of the measurement setup comprising a MSA400 Polytec LDV for the read out and electrostatic excitation method.
	The actuation voltage is generated by the LDV junction box and is amplified by a \SI{40}{\times} voltage amplifier that drives the levitating plate into resonance using electrostatic actuation. The electrostatic force is generated via two electrodes beneath the levitating plate. By focusing MSA laser beam on the plate, the vibration signal is detected,and the acquired velocity is transferred to a PC for frequency response analysis.
	}
	\label{fig: levitation system}
\end{figure} 


Pyrolytic graphite (from Magnetladen Seiler GmbH) is cut into square plates of different side length $L$ using a laser cutter after which their surfaces are slightly polished using fine sand paper (\SI{5}{\um} grain) to a thickness of 280 microns. The literature values of the material properties of the pyrolytic graphite resonator used for experiments and simulations are given in \cref{table: material properties of pg}. 
Our diamagnetic pyrolytic graphite plates are levitating above a checkboard arrangement of 4 cubic NdFeB ferromagnets (side length $D$=12 mm) with alternating magnetization (Fig. \ref{fig: levitation system}a). The remanent magnetic flux density of the cubic magnets is $B_r=\SI{1.4}{T}$. The magnetic field gradient in the $z$-direction provides an upward force on the plate to compensate the gravitational force. Together with the lateral magnetic gradients it provides a stable energy minimum that determines the plate's rest position. The figure shows that we define $x$ and $y$-axes along parallel to the plate edges. 

To probe the motion of the levitating resonator, we use the experimental setup shown in Fig. 1b. It includes an MSA400 Polytec LDV that measures the out-of-plane speed of the plate. In our experiments, the excitation voltage is generated by LDV junction box that drives the levitating plate into resonance using electrostatic forces. These forces are generated via two electrodes (thin copper tapes isolated from the magnets, not shown in Fig. 1a) beneath the levitating plate. When a voltage is applied between the two electrodes (Fig. 1b), the levitating plate acts as a floating electrode between the two electrodes, thus forming a capacitive divider. In the area at which the plate overlaps with the electrodes, an electrostatic downward force is exerted that depends on the overlapping area, voltage difference, and gap size. 
In order to efficiently excite different modes of the plate, an asymmetric arrangement of the electrodes is used, as shown in Fig. 1b. 
Due to this asymmetry, the electrostatic forces between each of the electrodes and the plate are different, and generate both a translational force and a rotational moment on the plate.
A periodic chirp voltage signal ($V_{AC}$=1.0 V)  is superposed on an offset voltage ($V_{DC}$=-1.2 V) and amplified by a voltage amplifier (40$\times$) that applies the voltage across the electrodes. The DC offset voltage is used to make sure that the electrostatic force, that is proportional to the square of the voltage, has a component of the same frequency as the chirp signal. With the LDV, the frequency response curves are measured, and by scanning the laser over the plate the mode shapes are obtained using the Polytec software.
The LDV measurements are conducted in a vacuum chamber over a pressure range of $10^{-4}$ to \SI{1000}{mbar}.
\begin{table}
	\centering
	\label{table: material properties of pg}
	\caption{Material properties of the levitating pyrolytic graphite.}
	\begin{tabular}{|c|c|c|c|}
		\hline 
		Property & Symbol & Value & Unit \\ 
		\hline 
		Density & $\rho$ & 2070 & $\si{kg/m^3}$ \\ 
		\hline 
		Susceptibility $\perp$ \cite{Simon2000}  & $\chi_z$ & -450 & $\times 10^{-6}$ \\ 
		\hline 
		Susceptibility $\parallel$ \cite{Simon2000} & $\chi_{x,y}$ & -85 & $\times 10^{-6}$ \\ 
		\hline 
		Conductivity $\perp$\cite{pappis1961properties} & $\sigma_z$ & 200 & S/m \\ 
		\hline 
		Conductivity $\parallel$\cite{pappis1961properties} & $\sigma_{x,y}$ & 200000 & S/m \\ 
		\hline  
	\end{tabular} 
\end{table}
\begin{figure*}
	\centering
	\subfigure{\label{fig:rigid body dynamics:a}\includegraphics[height=4.5cm]{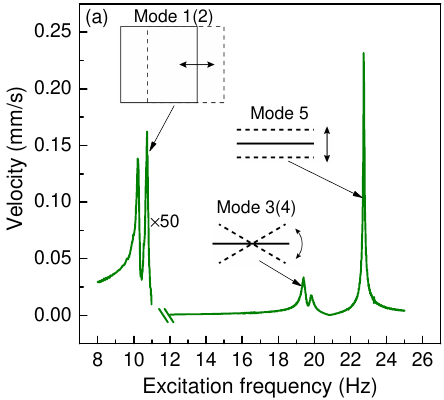}}
	\subfigure{\label{fig:rigid body dynamics:c}\includegraphics[height=4.5cm]{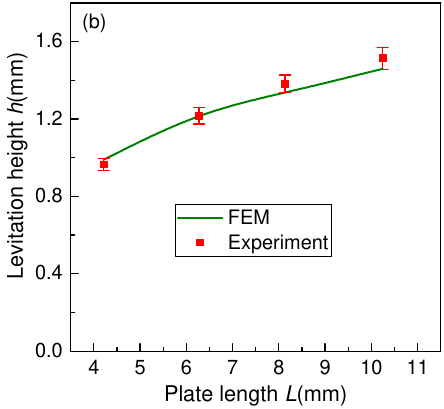}}
	\subfigure{\label{fig:rigid body dynamics:b}\includegraphics[height=4.5cm]{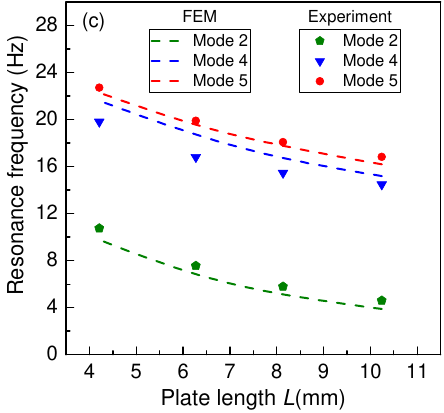}}
	\caption{Dynamic characterization of the levitating resonator in its rigid body modes. (a)  Frequency response curve of a $4.28\times4.14\times\SI{0.28}{mm^3}$ levitating graphite plate. The lowest frequency peaks have been multiplied by a factor 50 for visibility. Observed mode shapes are shown schematically near the corresponding resonance peaks. (b) Simulated and measured levitation height with different plate lengths (the thickness of all plates is 0.28mm). 
	(c) Experimental and simulated resonance frequencies at different plate lengths. All dimensions and material parameters used for the simulations are found in the text and in Table \ref{table: material properties of pg}, without free parameters.}
	\label{fig: frequency response}
\end{figure*}

Since there are 3 translational and 3 rotational degrees-of-freedom, the plate is expected to exhibit 6 different rigid-body resonances. In Fig. \ref{fig: frequency response}a, the levitating plate is driven from $\num{8}-\SI{25}{Hz}$, and 5 of these 6 rigid body modes are observed.  These mode shapes, as obtained by LDV and from camera movies, are schematically shown in Fig. \ref{fig: frequency response}a. Movies of the detected mode shapes are also shown in supplementary material  S1. The rotational mode around the $z$-axis is not observed, probably because it is not efficiently excited by the employed electrode configuration. Even if it were excited, it would not be efficiently detected by the LDV, that is only sensitive to motion in the $z$-direction. The frequency responses shown in Fig. \ref{fig: frequency response}a are determined at an off-centered point on the plate such that also the rotational modes 3 and 4 can be measured. Even though the amplitudes are small, it is surprising that modes 1 and 2 are detected using the vibrometer since their motion is in-plane, whereas the LDV is mainly sensitive to out-of-plane motion. 
It is likely that the in-plane motion of the plate is accompanied by an out-of-plane rotation or translation related to the shape of the magnetic potential well. We also note that both modes 1 and 2 as well as modes 3 and 4 are degenerate with almost identical resonance frequencies, and similar mode shapes (Fig. \ref{fig: frequency response}a). 

To investigate the plate size, side length $L$, dependence of the observed effects and for model verification, experiments are carried out on larger plates of 6.28, 8.14 and 10.25 mm. 
After placing the plates of different sizes above the magnet array, the levitation height of the plate is determined with a Keyence digital microscope (VHX-6000) using defocus method  (Fig. \ref{fig: frequency response}b). Using the LDV then the resonance frequencies of modes 2, 4 and 5 of the plate are determined and are shown in Fig. \ref{fig: frequency response}c. A clear reduction in resonance frequency with increasing plate size is observed. We will analyze these observations in more detail using quantitative models.


In the static levitating state only the gravitation force and magnetic force act on the plate. A model for the magnetic field generated by the 4 cubic NdFeB magnets is constructed and used to determine the levitation height. The FEM and analytical modelling results for the magnetic field \cite{furlani2001permanent,akoun19843d}, forces and levitation height, and its comparison to experiment can be found in the supplementary material S2.
The total magnetic force from the field on the plate $\mathbf{F_B}$ is determined as

\begin{equation}
    \label{FB}
    \mathbf{F_B}=\mathbf{\nabla} \int_\mathcal{V} \mathbf{M} \cdot \mathbf{B} \mathrm{d}\mathcal{V}=\frac{\mu_0}{2} \int_\mathcal{V} \mathbf{\nabla}(\chi_x H_x^2 + \chi_y H_y^2 +\chi_z H_z^2)\mathrm{d}\mathcal{V},
\end{equation}
where $\mathcal{V}$ is the volume of the plate, $H_{x,y,z}$ are the components of the magnetic field strength, $\mathbf{M}$ is the magnetization and $\mathbf{B}$ the magnetic flux density. In this analysis it is assumed that the plate does not significantly affect the magnetic field, since its relative magnetic permeability is close to 1.
The results of the FEM simulations, that determine the height at which the $z$-component of $\mathbf{F_B}$ is equal and opposite to the gravitational force, are shown in Fig. \ref{fig: frequency response}b. The dependence of the levitation height on plate size $L$ is well captured.

We also obtain the resonance frequencies of the plate numerically. The undriven free-body dynamics of translational modes 1,2 and 5 can be modelled by the differential equation $m\ddot{q}+c\dot{q}+kq=0$, where $q$ is the translational displacement and $m,c$ and $k$ are the mass, viscous damping coefficient and stiffness of that degree-of-freedom, respectively. Similarly for the rotational modes 3 and 4 the dynamics can be modelled by $I\ddot{\theta}+\Gamma\dot{\theta}+\mu \theta=0$, where $\theta$ is the rotational angle around the $x$ or $y$-axis and $I,\Gamma$ and $\mu$ are the moment of inertia, rotational viscous damping coefficient and torsional stiffness of the mode, respectively.

Since the gravitational force is independent from position, the stiffness of the resonant modes is solely determined by the derivative of the magnetic force $\mathbf{F_B}$, or magnetic torque, with respect to translational or rotational motion around its equilibrium position. When the plate translates or rotates, the spatial volume $\mathcal{V}$ over which the integral (\ref{FB}) is taken changes, and results in a restoring force or moment. For the torques, the integral in (\ref{FB}) is taken over the torque per volume element $\mathbf{r}\times \mathrm{d}\mathbf{F_B}$, where $\mathbf{r}$ is the position of the element with respect to the rotational axis. Using these integrals, the translational and torsional stiffnesses of the rigid body modes of the plate are numerically obtained.

From the FEM simulations the resonance frequencies are computed using $2 \pi f_{res}=\sqrt{k/m}$ and $2\pi f_{res}=\sqrt{\mu/I}$ for the translational and rotational modes, respectively. The FEM results are shown in Fig. \ref{fig: frequency response}c and show close agreement to the experimental results. The observed reduction in resonance frequency is less than what would be expected for a fixed stiffness system for which $f_{res}
\propto L^{-1}$. This indicates that the stiffnesses of the system increase with plate size despite the higher levitation height, which can be attributed to the larger volume over which the integral (\ref{FB}) is taken.


\begin{figure*}
	\centering
	\subfigure{\label{fig:rigid body dynamics:a}\includegraphics[height=4.5cm]{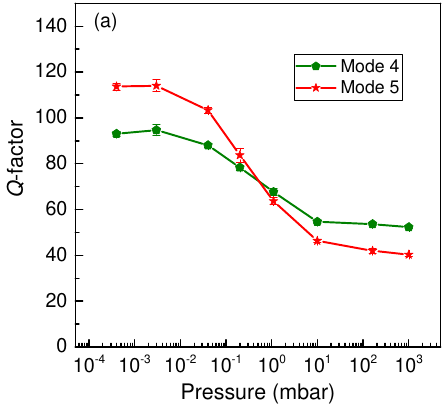}}
	\subfigure{\label{fig:rigid body dynamics:b}\includegraphics[height=4.5cm]{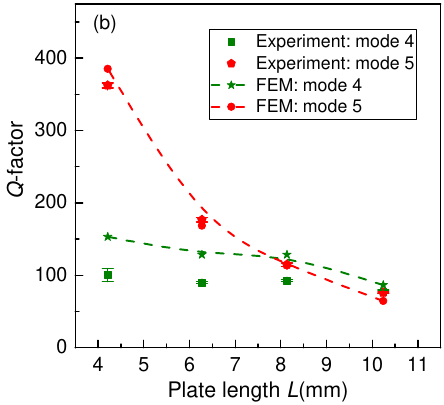}}\\
	\subfigure{\label{fig:rigid body dynamics:c}\includegraphics[height=4.5cm]{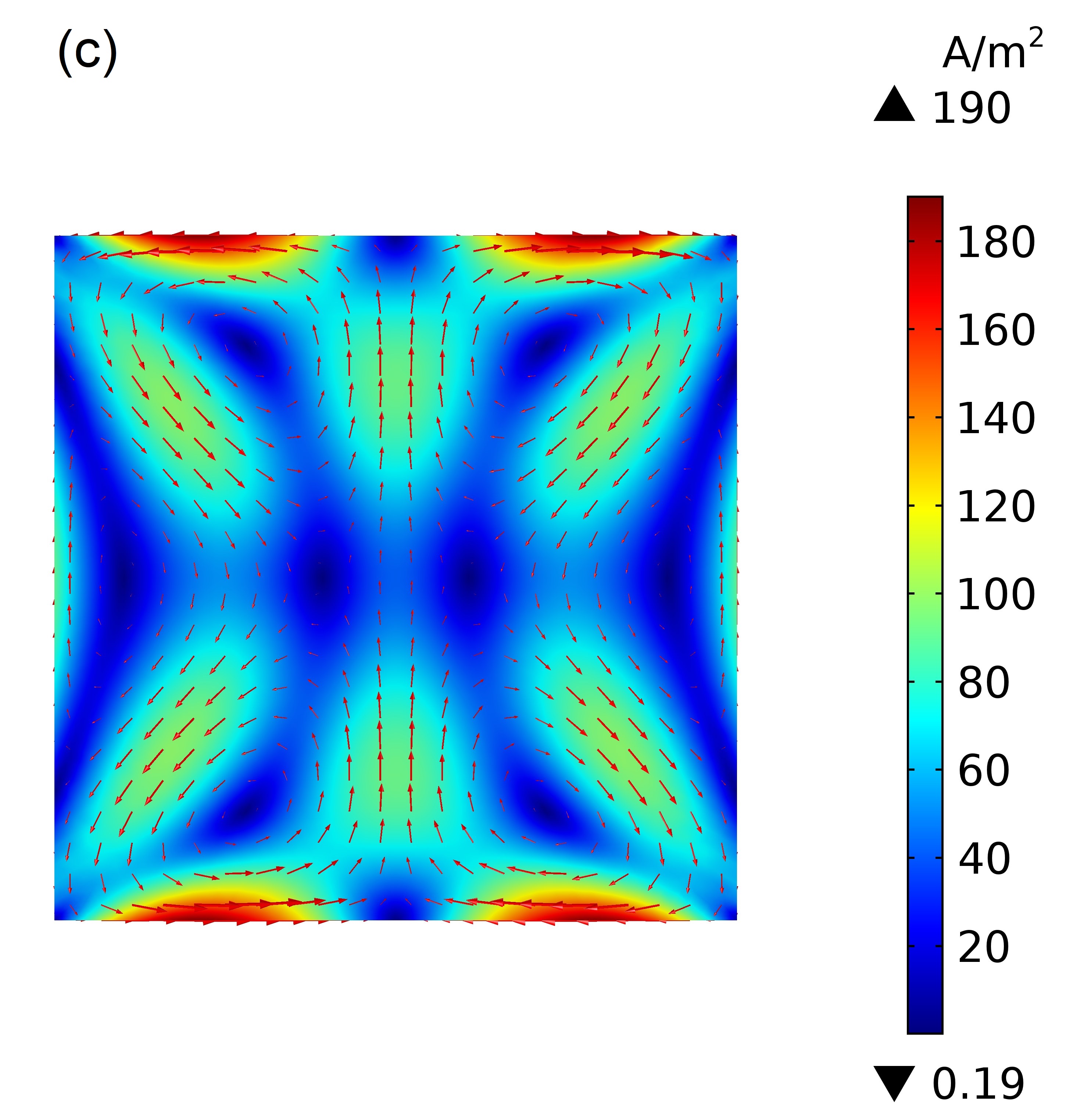}}
	\subfigure{\label{fig:rigid body dynamics:d}\includegraphics[height=4.5cm]{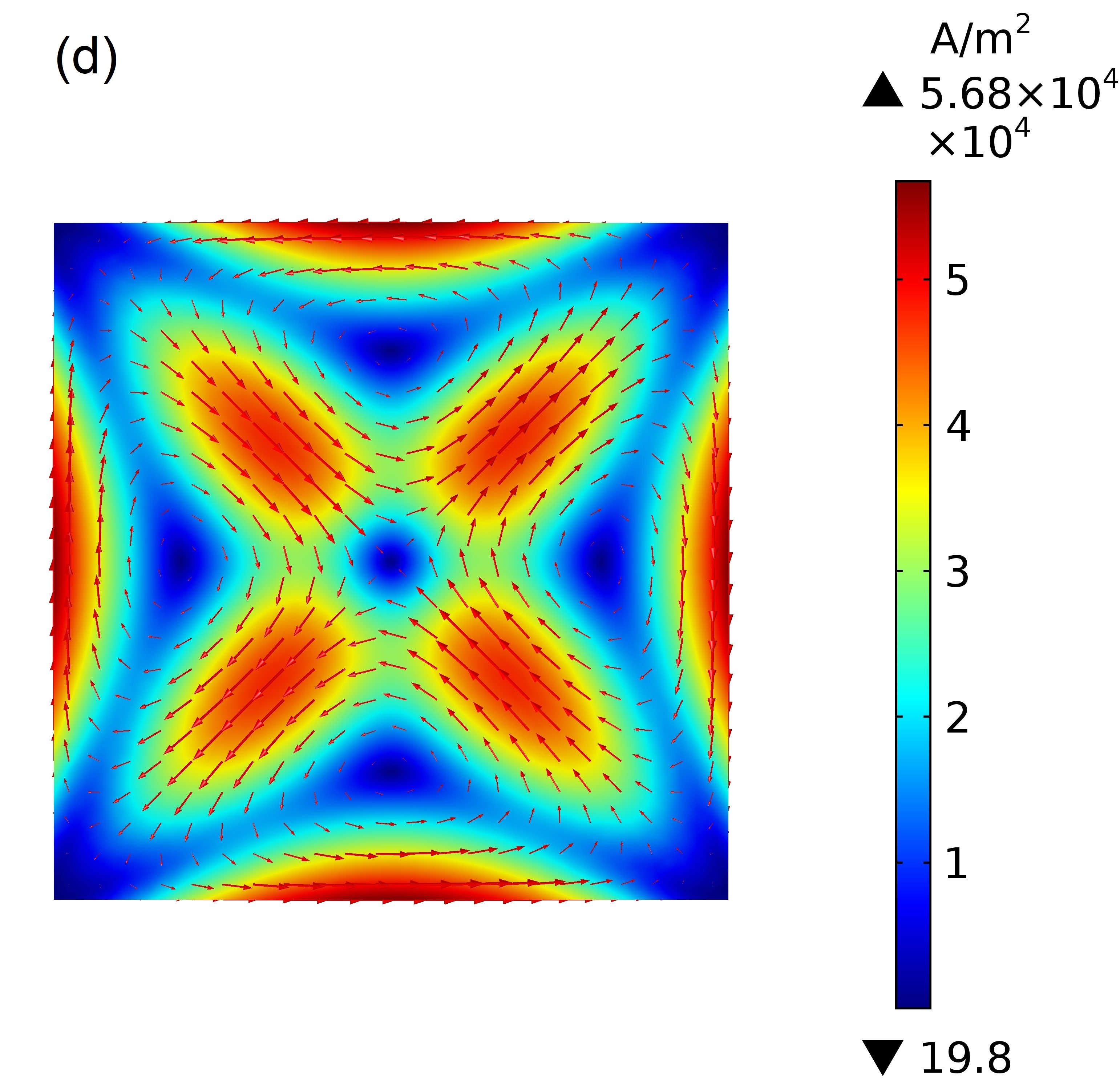}}
	\subfigure{\label{fig:rigid body dynamics:f}\includegraphics[height=4.5cm]{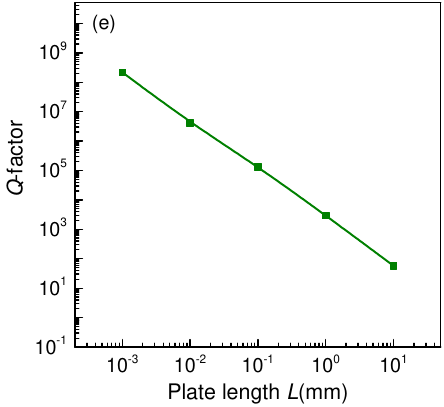}}
	\caption{Damping of levitating resonators. (a) Pressure-dependent $Q$ of a $8.03\times8.24\times\SI{0.28}{mm^3}$ levitating plate. (b) Experimental and simulated $Q$s of the levitating resonators of different plate lengths $L$ (thickness $t$=0.28mm). All measurements have been performed at room temperature and low pressure (\SI{0.001}{mbar}). (c) and (d) Eddy current density simulations for modes 4 and 5, respectively. The colour map indicates the eddy current density and the arrows show the trajectory of the eddy currents. (e) Simulation of the $Q$ as a function of plate length $L$ down to the microscale. In this simulation the plate thickness $t$ and magnet size $D$ is also scaled down ($D=1.2 L$ and $t=0.03 L$).}
	\label{fig: damping}
\end{figure*}

For the rigid body modes of diamagnetically levitating resonators, many types of dissipation are negligible. In particular, mechanical friction, radiation loss, surface and material damping are all expected to be negligible or absent during rigid body vibrations. Air damping is still significant, and can be eliminated by performing experiments at low pressure. In Fig. \ref{fig: damping}a it is shown that the resonator's $Q$ significantly increases and saturates at lower pressures. The saturation shows that at a pressure of $10^{-4}$ mbar, the damping of the rigid body modes is dominated by another mechanism, that is eddy current damping. To verify this, we compare the measured $Q$s on plates of different sizes in Fig. \ref{fig: damping}b. Included in the figure are also our FEM simulations of eddy current damping 

To quantify the eddy current damping of the levitating resonator, we have developed a FEM model to evaluate the eddy current damping force. When a conductor moves with velocity vector $\mathbf{v}$ through a magnetic flux density field $\mathbf{B}$, the charge carriers inside the conductor feel an electric field $\mathbf{v}\times \mathbf{B}$ due to the Lorentz' force in addition to the field from the electric potential $V_e$, that generates an eddy current density $\mathbf{J}$ given by
\begin{equation}
	\mathbf{J}=-\mathbf{\sigma}\nabla V_e+\mathbf{\sigma}(\mathbf{v}\times \mathbf{B}),
	\label{eq: current density}
\end{equation}
where $\mathbf{\sigma}$ is the electrical conductivity matrix, that has nonzero diagonal elements $\sigma_{x}$, $\sigma_{y}$, and $\sigma_{z}$. 
By combining Eq.\ref{eq: current density} with the current continuity condition $\nabla\cdot\mathbf{J}=0$ and the boundary condition $\mathbf{J}\cdot\mathbf{n}=0$ ($\mathbf{n}$ is the unit vector perpendicular to the boundary), the eddy current density distribution $\mathbf{J}$ and potential $V_e$ can be determined numerically for known $\mathbf{v}$, $\mathbf{\sigma}$, $\mathbf{B}$ and plate dimensions. The simulated current density $\mathbf{J}$ distribution through the plate for the rotational mode 4 and translational mode 5 are shown in Figs. \ref{fig: damping}c,d respectively.

The total damping contribution due to eddy currents can now be evaluated for the translational modes by integrating the eddy current forces $\mathbf{J}\times \mathbf{B}$ to obtain $\mathbf{F}_{\rm eddy}$ \cite{ebrahimi2008design}. Similarly for the rotational modes the torques $\mathbf{r}\times (\mathbf{J}\times \mathbf{B})$ can be integrated to obtain the total torque $\mathbf{\tau}_{\rm eddy}$. Since for thin plates and motion under consideration, the eddy currents run mainly in-plane and the motion is out-of-plane, it is the in-plane component of $\mathbf{B}$ that contributes mainly to the relevant out-of-plane force $\mathbf{F}_{\rm eddy} \propto (\mathbf{v} \times \mathbf{B}) \times \mathbf{B}$. Since the in-plane component of the $\mathbf{B}$ field is largest near the boundary between the magnets, it is expected that these regions contribute most to the $\mathbf{F}_{\rm eddy}$.

As expected, the eddy current damping force $\mathbf{F}_{\rm eddy}$ is found to be proportional and in the opposite direction of the velocity $\dot{q}$ and similarly the torque $\mathbf{\tau}_{\rm eddy}$ is opposite and proportional to the rotational velocity $\dot{\theta}$. From the proportionality constants the coefficients $c$ and $\Gamma$ for the translational and rotational modes can be determined. Using our FEM simulations, and material parameters given in table \ref{table: material properties of pg}, we found the $Q$s associated with modes 4 and 5 for different plate lengths $L$ as follows
\begin{equation}
    \label{eq: Q}
    Q=\frac{2\pi m f_\mathrm{res}}{c}, ~~ Q = \frac{2\pi I f_\mathrm{res}}{\Gamma}.
\end{equation}

The simulated $Q$ corresponds well to the experimental values as shown in Fig. \ref{fig: damping}b. This provides evidence that eddy current damping can account largely for the observed $Q$s, and their size dependence for the rigid body modes of diamagnetically levitating plates. 
A difference is observed between the simulated and measured $Q$ of mode 4, which is attributed to the fact that the actual experimentally obtained rotation is observed to occur not exactly around the $x$ or $y$-axis.
Fig. \ref{fig: damping}b shows a steep increase in $Q$ with decreasing plate length $L$ for translational mode 5. A reduction of plate length $L$ by a factor of 2 results in an increase of $Q$ by a factor of $\sim$4. For the rotational mode 4 such an increase in $Q$ is not observed.
The observed experimental trend suggests that very high $Q$s might be achieved in diamagnetically levitating plates of microscopic dimensions.
To test this hypothesis we simulate the $Q$ of the $z$-direction translational mode 5 for levitating graphite plate resonators with $L$=10$^{-4}$-10$^{1}$ mm. Besides scaling the lateral dimensions of the plate, also the dimension of the magnets $D=1.2 L$ and plate thickness $t=0.03 L$ are scaled proportionally. For each value of $L$, first the magnetic field and levitation height are calculated, then the resonance frequency, eddy currents and $Q$ are determined according to the procedure outlined before. The result is plotted in Fig. \ref{fig: damping}e.
For a reduction of $L$ by a factor 10$^4$ the $Q$ increases by a factor $3.8\times10^6$. For plate sizes of the order of 1 $\mu$m, $Q$s above 100 million might be achievable, competitive to the $Q$ of the best mechanical resonators currently available \cite{norte2016mechanical}. This suggests that levitating  nano/micro particles could be interesting candidates for realizing high-Q resonators for accurate sensors and for studying quantum mechanics at room temperature \cite{hsu2016cooling,o2019magneto,lewandowski2020high, norte2016mechanical}.

In conclusion, we experimentally study the rigid body motion of diamagnetically levitating resonators at different pressures. By levitating graphite plates above magnets we eliminate external mechanical effects, such that their dynamics is solely governed by magnetic field. The levitation height, resonance frequencies and $Q$s are measured as a function of plate size by laser Doppler vibrometry, and are modeled effectively using FEM simulations. In particular the $Q$ of the out-of-plane translational mode is found to increase significantly with reducing dimensions. Using simulations evidence is provided that this increase in $Q$ continues at smaller dimensions where $Q$ factors above 100 million might be attainable, making levitating diamagnetic resonators an interesting candidate for high-$Q$, low-noise oscillators and sensors.

See the supplementary material for (S1) movies of the rigid body modes of a diamagnetically levitating resonator; (S2) simulation of the magnetic levitation height.

This work was carried out under the 17FUN05 PhotOQuanT project, which has received funding from the EMPIR program, co-financed by the Participating States and the European Union’s Horizon 2020 research and innovation program. This work has also received funding from ERC starting grant ENIGMA (802093) and Graphene Flagship (881603). X.C acknowledges financial support from China Scholarship Council.


The data that support the findings of this study are available from the corresponding author upon request.

\bibliography{ref-levitating-plate}
\pagebreak
\onecolumngrid

\section*{Supplementary Materials}
\subsection*{S1: Movies of the rigid body modes of a diamagnetically levitating resonator} As supporting information, movies of the rigid body motion of a levitating plate are included. The slow-motion movies are recorded at a frame rate of 240 fps and played back eight times slower. They are recorded on a $4.28\times4.14\times \SI{0.28}{mm}$ pyrolytic graphite plate that levitates
above 4 permanent cubic NdFeB magnets, with an edge length of \SI{12}{mm}, and alternating out-of-plane magnetization.
The plate is actuated by electrostatic force using two asymmetric electrodes (copper tapes) that are isolated from the magnets. 
A sinusoidal voltage signal ($V_{AC}$=1.0 V) at different specific frequency ($10.23, 10.75, 19.41, 19.81$ and $\SI{22.72}{Hz}$) is superposed on an offset voltage ($V_{DC}$=-1.2 V) and amplified by a voltage amplifier (40$\times$) that applies the voltage across the electrodes.
\subsection*{S2: Simulation of the magnetic levitation height}

In order to determine the levitation height, we calculate the magnetic field both using an analytical derivation and using the Finite Element Method (FEM). The FEM simulations are performed using \emph{COMSOL Multiphysics 5.3a}. For comparison with experiments we model the field distribution of 4 permanent magnets of 12$\times$12$\times 12$ mm$^3$ with a remanent magnetic flux density $B_r$=1.4 T and rounded edges with a fillet radius of \SI{1}{mm}. The summed magnetic field $\mathbf{B}(x, y, z)$ of the magnets is determined analytically and using FEM. Then, assuming that the influence of the diamagnetic plate on the field is negligible, the integrated magnetic force on the diamagnetic plate, $\mathbf{F_B}$, can be determined using
\begin{equation}
    \setcounter{equation}{1}
    \label{FB}
    \renewcommand{\theequation}{S\arabic{equation}}
    \mathbf{F_B}=\boldsymbol{\nabla} \int_\mathcal{V} \mathbf{M} \cdot \mathbf{B} \mathrm{d}\mathcal{V}=\frac{\mu_0}{2} \int_\mathcal{V} \boldsymbol{\nabla}(\chi_x H_x^2 + \chi_y H_y^2 +\chi_z H_z^2)\mathrm{d}\mathcal{V},
\end{equation}
where $\chi_x, \chi_y, \chi_z$ are the magnetic susceptibility of the levitating plate in $x, y, z$ directions, $\mathcal{V}$ is the volume of the plate, and $\mathbf{M}$ is the plate's magnetization. The components of the magnetic field inside the plate are $H_{x,y,z}=B_{x,y,z}/\mu$, where the magnetic permeability $\mu \approx \mu_0$ for graphite. 

\setcounter{figure}{0}
\begin{figure}[H]
   \centering
   \renewcommand{\thefigure}{S\arabic{figure}}
    \includegraphics[width=6cm]{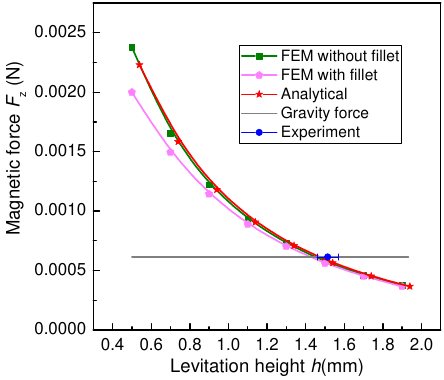}
    \caption{Magnetic force as a function of levitation height calculated by analytical and FEM modelling for a graphite plate of $10.23\times10.26\times\SI{0.28}{mm^3}$ above 4 laternating NdFeB magnets of  $12\times12\times\SI{12}{mm^3}$. The levitation height $h$ is defined as the distance from the top surface of the levitating plate to the top surface of magnets.}
    \label{fig:F-H}
\end{figure}

The magnetic field is determined and  in Fig. \ref{fig:F-H} we compare the volume integrated upward magnetic force $F_{B,z}$ calculated analytically for 4 cubes  without fillets to the force obtained from FEM simulations as a function of levitation height $h$. The correspondence between analytical and FEM simulation verifies the accuracy of the FEM simulation. In the same figure the integrated downward gravitational force on the diamagnetic plate is shown. The plate levitates at the height $h$ where the curves intersect. This agrees well with the measured levitation height (solid blue circle) of the plate with dimensions $10.23\times10.26\times\SI{0.28}{mm^3}$.

\end{document}